\documentclass[twocolumn,amsmath,amssymb,prb,aps,superscriptaddress]{revtex4-1}
\usepackage{txfonts}
\usepackage{pifont}
\usepackage{amssymb}
\usepackage{dcolumn}
\usepackage{amsmath}
\usepackage[dvips]{epsfig}
\usepackage{array}
\usepackage{color}
\usepackage{ulem}
\usepackage{subfig}
\usepackage{caption}
\usepackage{graphicx}
\usepackage{titlesec}
\usepackage{floatrow}
\usepackage[colorlinks=true,linkcolor=blue,citecolor=blue,urlcolor=blue,]{hyperref}
\usepackage[T1]{fontenc}

\newcolumntype{C}[1]{>{\centering}p{#1}}
\makeatletter
\def\btt#1{\texttt{\@backslashchar#1}}%
\DeclareRobustCommand\bblash{\btt{\@backslashchar}}%
\makeatother

\begin{document}
\title{Antiferromagnetic two-dimensional transition-metal nitride Co$_2$N$_2$ layer with high Neel temperature and Dirac fermions}

\author{Lujia Tian}
\affiliation{College of Physics and Engineering, Qufu Normal University, Qufu, Shandong 273165, China}
\author{Lihui Han}
\affiliation{College of Physics and Engineering, Qufu Normal University, Qufu, Shandong 273165, China}
\author{Yuanfang Yue}
\affiliation{College of Physics and Engineering, Qufu Normal University, Qufu, Shandong 273165, China}
\author{Huazhen Li}
\affiliation{School of Physics, Renmin University of China, Beijing 100872, China}
\affiliation{Key Laboratory of Quantum State Construction and Manipulation (Ministry of Education), Renmin University of China, Beijing 100872, China}
\author{Z. Y. Xie}
\email{qingtaoxie@ruc.edu.cn}
\affiliation{School of Physics, Renmin University of China, Beijing 100872, China}
\affiliation{Key Laboratory of Quantum State Construction and Manipulation (Ministry of Education), Renmin University of China, Beijing 100872, China}
\author{Xun-Wang Yan}
\email{yanxunwang@163.com}
\affiliation{College of Physics and Engineering, Qufu Normal University, Qufu, Shandong 273165, China}

\date{\today}

\begin{abstract}
Two-dimensional (2D) transition metal nitrides have a wide prospect of applications in the fields of physics, chemistry, materials, etc. However, 2D transition metal nitrides with strong magnetism, especially high N$\rm \acute{e}$el temperature, are very scarce. Based on the first-principles calculations within the framework of density functional theory, we design two 2D transition-metal nitrides \textit{M}$_2$N$_2$ (\textit{M} = Ti, Co), in which the transition metal atoms and the N atoms form a 2D layer with a wrinkled structure. The structural stability is demonstrated by the cohesive energy, formation energy, elastic constants, phonon spectra and molecular dynamics simulations. Elastic moduli calculations reveal that the mechanical properties of the two structures are anisotropic. Spin-polarized calculations show that Ti$_2$N$_2$ is a 2D ferromagnetic material while Co$_2$N$_2$ is a 2D antiferromagnetic semimetal with a Dirac point at Fermi level. Furthermore, by solveing the Heisenberg model by Monte Carlo method, we discover that the 2D Co$_2$N$_2$ layer is a high-temperature antiferromagnetic material and the N$\rm \acute{e}$el temperature is up to 474 K.
Therefore, our findings provide a rare antiferromagnetic 2D material with both high critical temperature and Dirac Fermions.

\end{abstract}

\maketitle
\section{INTRODUCTION}
After the discovery of graphene in 2004 \cite{Novoselov2004}, boron nitride \cite{Song2010}, silicene \cite{Lalmi2010}, borophene \cite{Mannix2015}, stanene \cite{Saxena2016}, transition metal dichalcogenides \cite{Coleman2011}, and MXenes \cite{Naguib2011} have been fabricated in experiments. These 2D materials possess atomic-level thickness and their carrier migration and thermal diffusion are confined within the 2D space, a variety of unique mechanical, thermal, electronic, magnetic, and optical properties are usually found in them, which lead to great potential applications in many fields \cite{Tan2017}. Hence, 2D materials attract considerable interest among researchers and become the hot research topic in the area of physics and materials.
Compared to metal oxides and chalcogenides, bulk transition metal nitrides typically exhibit high hardness and high melting points, which are associated with applications such as refractory materials, cutting tools, and hard coatings \cite{Salamat2013}.
For two-dimensional transition metal nitrides, what are the excellent physical properties is a topic worthy of expectation.

The Mermin-Wagner theorem states that at any nonzero temperature, a one- or two-dimensional isotropic spin Heisenberg model with finite-range exchange interaction can be neither ferromagnetic nor antiferromagnetic \cite{Mermin1966},
which leads to a misunderstanding that it is impossible to fabricate magnetic two-dimensional materials in experiments. The misunderstanding hinders the development of magnetic 2D materials.
In fact, the long-range magnetic order can exist in a 2D material when the isotropy of magnetic exchange or spin orientation is broken.
Until 2017, the first magnetic 2D material, ferromagnetic CrI$_3$ monolayer \cite{Huang2017}, was synthesized in experiments.
In recent years, several ferromagnetic 2D materials were discovered, such as Cr$_2$Ge$_2$Te$_6$ \cite{Gong2017}, VSe$_2$ \cite{Bonilla2018}, Cr$_2$Te$_3$ \cite{Zhong2023}, Fe$_3$GaTe$_2$ \cite{Zhang2022b}, while only a few antiferromagnetic 2D materials such as CrPS$_4$ \cite{Lee2017} and FePS$_3$ \cite{Kargar2020} are synthesized.

The next generation microelectronic devices must meet the demand for ultra-high density and computing speed, as well as ultra-low energy consumption.
Antiferromagnetic 2D materials have great potential in this regard.
Firstly, the neighboring magnetic moments are arranged in the opposite direction in the antiferromagnetic materials. The net magnetic moment sums to zero, rendering the magnetism of antiferromagnets imperceptible externally. The microscopic devices made by antiferromagentic materials are insensitive to disturbing magnetic fields, and the memory cell fabricated from antiferromagnets would not affect magnetically their neighbors, conducive to higher density of storage units \cite{Zelezny2018}.
Secondly, the resonant frequency of antiferromagnetic materials is the THz range due to the inter-spin-sublattice exchange, which shifts the accessible electrical writing speeds up by three orders of magnitude compared to ferromagnets \cite{Jungwirth2018}.
Therefore, recent research in 2D spintronics has increasingly focused on antiferromagnetic (AFM) materials due to the robustness against magnetic disturbances and their fast dynamic behaviors \cite{Wadley2016,Baltz2018,Wang2023b,Han2023}.

Dirac points in the band structure of 2D materials have been a fascinating subject of research. Around the Dirac points, two bands cross with linear dispersion, where the low-energy electrons behave like relativistic massless Dirac fermions distinct from the usual Schr$\rm {\ddot o}$dinger fermions.
 The magnetic ordering breaks the time-reversal symmetry and hence may destroy the Dirac points. Therefore, 2D Dirac materials are usually nonmagnetic and AFM 2D Dirac materials are scarce. In a magnetic crystal cell, although the spatial symmetry (P) and the time reversal symmetry (T) are
individually broken by the magnetic order, the Dirac point still can emerge if the combined PT symmetry is preserved \cite{Li2019c,Young2017}.
 Electronic transport in antiferromagnetic materials is affected by spin scattering, which significantly decrease the speed and efficiency of the device. To tackle the issue, the integration of massless Dirac Fermions within 2D antiferromagnetic structures has been proven to be a promising strategy \cite{Li2019c}.
This renders the antiferromagnetic 2D materials with Dirac points of great significance for the emerging spintronic devices.

High N$\rm \acute{e}$el temperature is a key parameter in practical applications for an antiferromagnetic 2D material.
Despite the growing researches on 2D antiferromagnetic materials, only a few materials
have been reported, and most of these materials have low N$\rm \acute{e}$el temperature, which brings a great limit to the practical applications and hinders the development of spintronics. Therefore, it is an urgent topic to explore new 2D antiferromagnetic materials with both high N$\rm \acute{e}$el temperature and Dirac point.


In the work, we design a 2D transition metal nitride \textit{M}$_2$N$_2$ (\textit{M} = Ti, Co) by arranging transition-metal atoms and N atoms alternately to consist of an atom-level-thick sheet. Ti$_2$N$_2$ is a 2D ferromagnetic material while Co$_2$N$_2$ is a 2D antiferromagnetic material. Only one Dirac point emerges at Fermi level in the band structure of Co$_2$N$_2$ monolayer, leading to a semimetal character. In particular, the N$\rm \acute{e}$el temperature is up to 474 K in Co$_2$N$_2$, which makes it a high-temperature 2D antiferromagnet.

\section{COMPUTATIONAL DETAILS}

The calculations in this paper are performed by using the Vienna ab initio simulation package (VASP) program~\cite{Kresse1993,Kresse1996}. The generalized gradient approximation (GGA) with Perdew-Burke-Ernzerhof (PBE) formula~\cite{Perdew1996} as well as the projector augmented-wave method (PAW)~\cite{Bloch1994} for ionic potential are employed. The GGA + \textit{U} method is applied to consider the correction of electron correlation~\cite{Dudarev1998}. The plane wave basis cutoff is set to 600 eV, and the convergence thresholds for the total energy and force are ${10}^{-5}$ eV and 0.01 eV/\AA~ used in structural optimization. We adopt a mesh of 26 $\times$ 26 $\times$ 1 k points for the Brillouin zone integration, and a vacuum layer of 15 \AA~ is taken in the z-direction to simulate a single layer of \textit{M}$_2$N$_2$. The ab initio molecular dynamics (AIMD) simulation is used to examine the thermal stability. A canonical ensemble (NVT) system is used with the temperature held at 1000 K for 5 ps with a time step of 1 fs~\cite{Martyna1992}. The temperature of the phase transition in the \textit{M}$_2$N$_2$ system is evaluated by solving the Heisenberg model combined with Monte Carlo method, and the 200 $\times$ 200 $\times$ 1 lattice is used in the Monte Carlo simulation.

\section{RESULTS AND DISCUSSION}
\subsection{Atomic structure}

\captionsetup[figure]{name={Fig.},labelsep=period}
\begin{figure}[ht]
\centering
\includegraphics[scale=0.35]{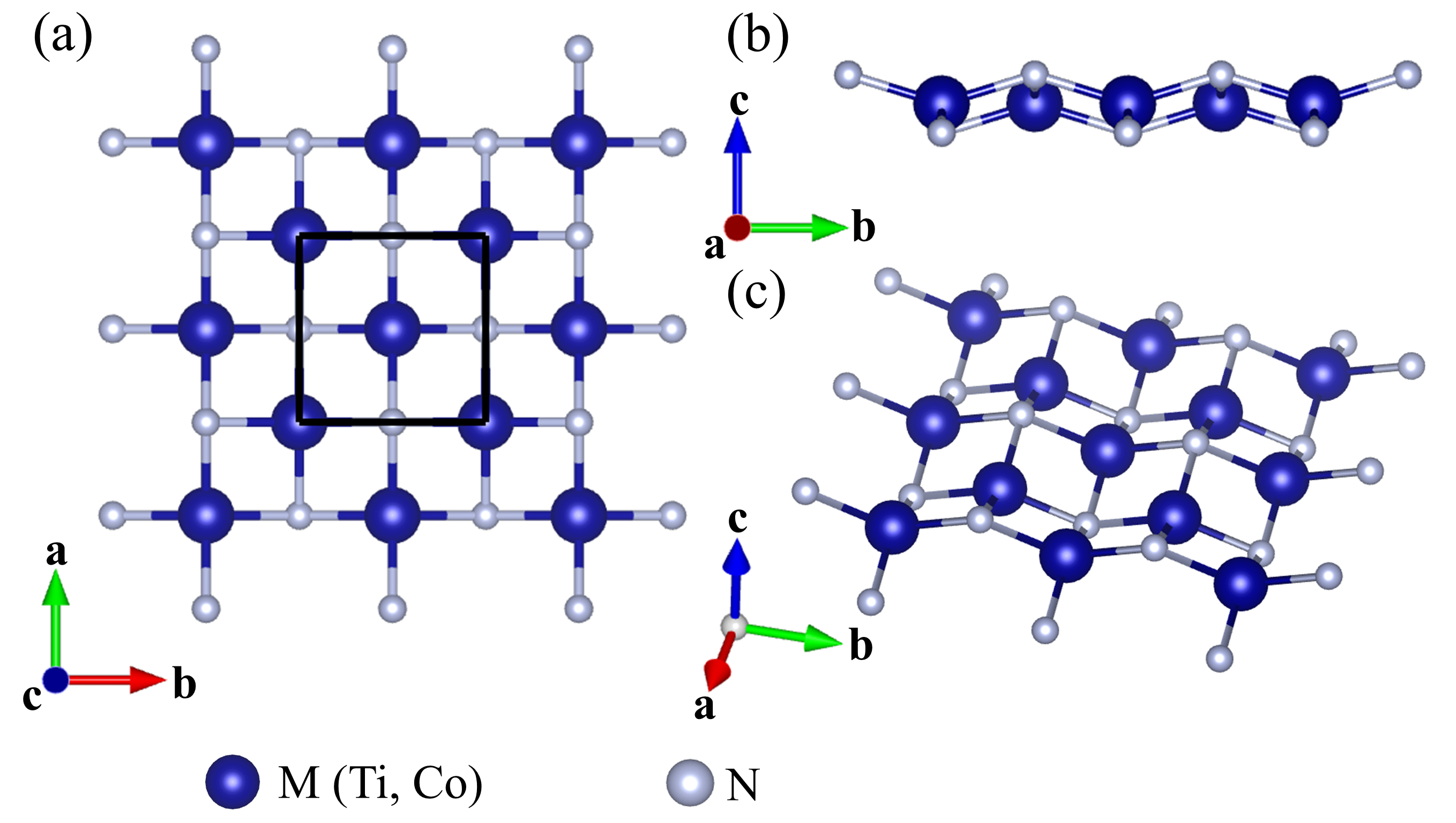}
\caption{Atomic structure of \textit{M}$_2$N$_2$ (\textit{M} = Ti, Co). (a) Top view, (b) Side view, (c) Oblique view. The unit cell is marked with a solid black line square. }
\label{fig.1}
\end{figure}

The atomic structure of \textit{M}$_2$N$_2$ (\textit{M} = Ti, Co) is shown in Fig. \ref{fig.1}. The primitive cell of \textit{M}$_2$N$_2$ consists of two \textit{M} atoms and two N atoms, which are arranged alternately to form an atom-level-thick sheet. The two atoms are represented by dark blue and silver balls, where an \textit{M} atom is tetra-coordinated by the N atoms. The symmetry of \textit{M}$_2$N$_2$ belongs to the space group P4/nmm (No.129) and is subordinate to the $D_{4h}$ point group. The optimized lattice parameters $a$ of the \textit{M}$_2$N$_2$ (\textit{M} = Ti, Co) structures are 3.84 \AA~ and 3.29 \AA, respectively. The height difference $\Delta$z between the two kinds of N atoms at the top and bottom positions is 1.19 \AA~ for Ti$_2$N$_2$ and 1.83 \AA~ for Co$_2$N$_2$.
From the results, the lattice parameter decreases with reducing atomic radius from Ti to Co, while the thickness of 2D layer increases with reducing atomic radius.

\subsection{Structural stability}
In order to determine the stability of the two 2D materials, we perform the calculations of the cohesive energy, formation energy, elastic constants, phonon spectra, and molecular dynamics simulation to verify that \textit{M}$_2$N$_2$ (\textit{M} = Ti, Co) has good mechanical, dynamical, and thermal stability.

\subsubsection*{\rm\textbf{1.\quad Cohesive energy and Formation energy}}
We first calculate the cohesive energy of \textit{M}$_2$N$_2$ (\textit{M} = Ti, Co), which is defined as
\begin{equation}
    {E}_{coh}=\frac{1}{4}({E}_{tot}-2{E}_{metal}-2{E}_{N})
\end{equation}
where $E_{tot}$ is the total energy of a formula unit cell, $E_{metal}$ and $E_{N}$ are the single atom energy of different elements. We obtain the cohesive energies of -6.34 eV/atom and -4.88 eV/atom, similar to the cohesive energies of the silicene and MoS$_2$ monolayer, -3.96 eV/atom and -5.14 eV/atom. From the view of cohesive energy, \textit{M}$_2$N$_2$ (\textit{M} = Ti, Co) is stable.

Next, we perform the formation energy calculations. The formation energy $E_{form}$ is defined as
\begin{equation}
    {E}_{form}=\frac{1}{4}({E}_{tot}-2{E}_{metal}-{E}_{{N}_{2}})
\end{equation}
in which $E_{tot}$, $E_{metal}$ and  $E_{N_{2}}$ are the total energy, bulk metal energy per atom, and the energy of a nitrogen molecule. The formation energies of \textit{M}$_2$N$_2$ (\textit{M} = Ti, Co) are -1.08 eV/atom and 0.38 eV/atom. Nitrogen molecules typically have very low energies due to the existence of N$\equiv$N, so metal nitrides often possess positive formation energy. From Ti to Co, as electronegativity increases, it becomes increasingly difficult for electrons to be transferred from the metal atoms to the N atoms. Consequently, bonding becomes more and more difficult and the formation energy from Ti$_2$N$_2$ to Co$_2$N$_2$ becomes higher. For comparison, we perform the formation energy calculations for the nitrides CuN$_3$~\cite{Wilsdorf1948}, PtN$_2$~\cite{Crowhurst2006}, and g-C$_3$N$_4$~\cite{Groenewolt2005}, which have already been synthesized in experiments. The formation energies of them are 0.55 eV/atom, 0.42 eV/atom, and 0.35 eV/atom, respectively. So, the formation energies of Ti$_2$N$_2$ and Co$_2$N$_2$ are comparable to them, indicating that there is large possibility to fabricate the Ti$_2$N$_2$ and Co$_2$N$_2$ in experiments.

\captionsetup[figure]{name={Fig.},labelsep=period}
\begin{figure}[htbp] \centering
\includegraphics[width=8.0cm]{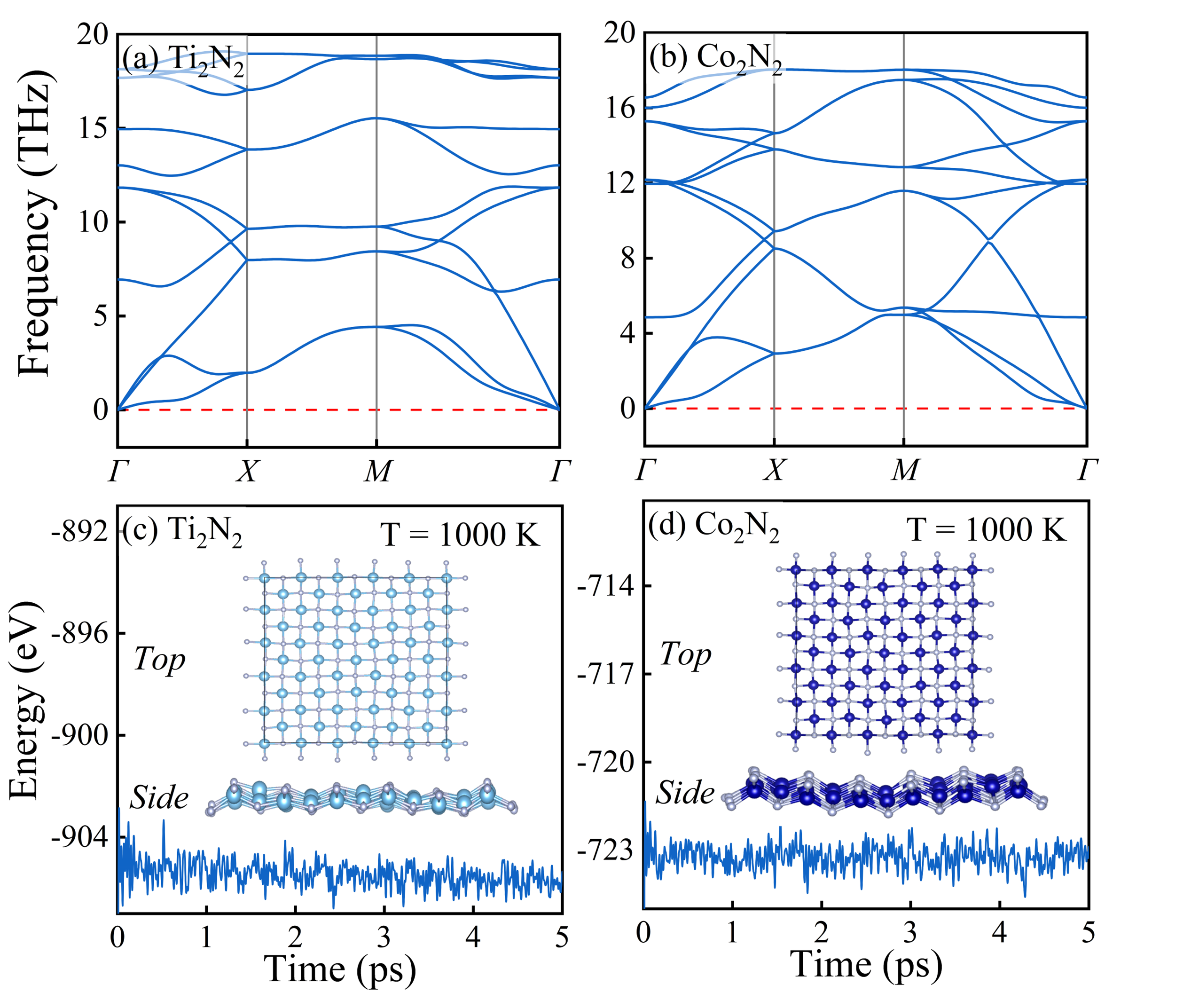}
\caption{(a)-(b) Phonon spectra of Ti$_2$N$_2$ and Co$_2$N$_2$. (c)-(d) The evolution of total potential energy with the time at 1000 K for 5 ps, where the insets are the top and side views of the final configuration of Ti$_2$N$_2$ and Co$_2$N$_2$.}
\label{fig.2}
\end{figure}

\subsubsection*{\rm\textbf{2.\quad Phonon spectra and Molecular dynamic simulation}}

Based on the DFTP method, we compute the phonon spectra of \textit{M}$_2$N$_2$ (\textit{M} = Ti, Co) to check its dynamical stability. The phonon spectra of \textit{M}$_2$N$_2$ (\textit{M} = Ti, Co) is shown in Fig. \ref{fig.2}(a)$\sim$(b). The highly symmetric points in reciprocal space are $\mathit{\Gamma}$(0 0 0), \textit{X}(0.5 0 0),  and \textit{M}(0.5 0.5 0). There is no imaginary frequency appearing, indicating that \textit{M}$_2$N$_2$ is dynamically stable.

After that, we further examine the thermal stability of \textit{M}$_2$N$_2$ (\textit{M} = Ti, Co) through first-principles molecular dynamic simulation. Fig. \ref{fig.2}(c)$\sim$(d) show the evolution of the total potential energy in \textit{M}$_2$N$_2$ with the time at 1000 K temperature for 5 ps. It can be seen that the total potential energy of the \textit{M}$_2$N$_2$ fluctuates around a certain value without a significant decrease, and the \textit{M}$_2$N$_2$ is able to maintain the original framework without broken bonds. Hence, the \textit{M}$_2$N$_2$ has good thermal stability.

\subsubsection*{\rm\textbf{3.\quad Elastic constants}}

\begin{table}[tbp] \centering
\caption{Elastic constants of \textit{M}$_2$N$_2$ (\textit{M} = Ti, Co). The unit is N/m.}
\renewcommand\tabcolsep{8.5pt}
\renewcommand\arraystretch{1.5}
\begin{tabular*}{8.5cm}{cccccc} \hline\hline
      & ${C}_{11}$   & ${C}_{22}$   & ${C}_{12}$  &  ${C}_{21}$ &  ${C}_{66}$   \\ \hline
Ti$_2$N$_2$ & 100.81 & 100.81 & 68.49 & 68.49 & 61.26  \\
Co$_2$N$_2$ & 121.85 & 121.85 & 79.23 & 79.23 & 102.79 \\ \hline\hline
\end{tabular*}
\label{Table.1}
\end{table}

In order to further investigate the mechanical stability of \textit{M}$_2$N$_2$ (\textit{M} = Ti, Co), we carry out calculations of elastic constants. The four nonzero 2D elastic constants for square lattice are ${C}_{11}$, ${C}_{22}$, ${C}_{12}$, and ${C}_{66}$. Meanwhile, due to symmetry, the square structures have the additional relation $C_{11}=C_{22}$~\cite{Liu2017}. The elastic constants calculated are shown in Table \ref{Table.1}. These elastic constants satisfy the two inequalities ${C}_{11}{C}_{22}-{C}_{12}{C}_{21}>0$ and $C_{66}>0$. Namely, they satisfy the mechanical stability Born criteria~\cite{Born1954}.

\captionsetup[figure]{name={Fig.},labelsep=period}
\begin{figure}[ht]
\centering
\includegraphics[width=7.5cm]{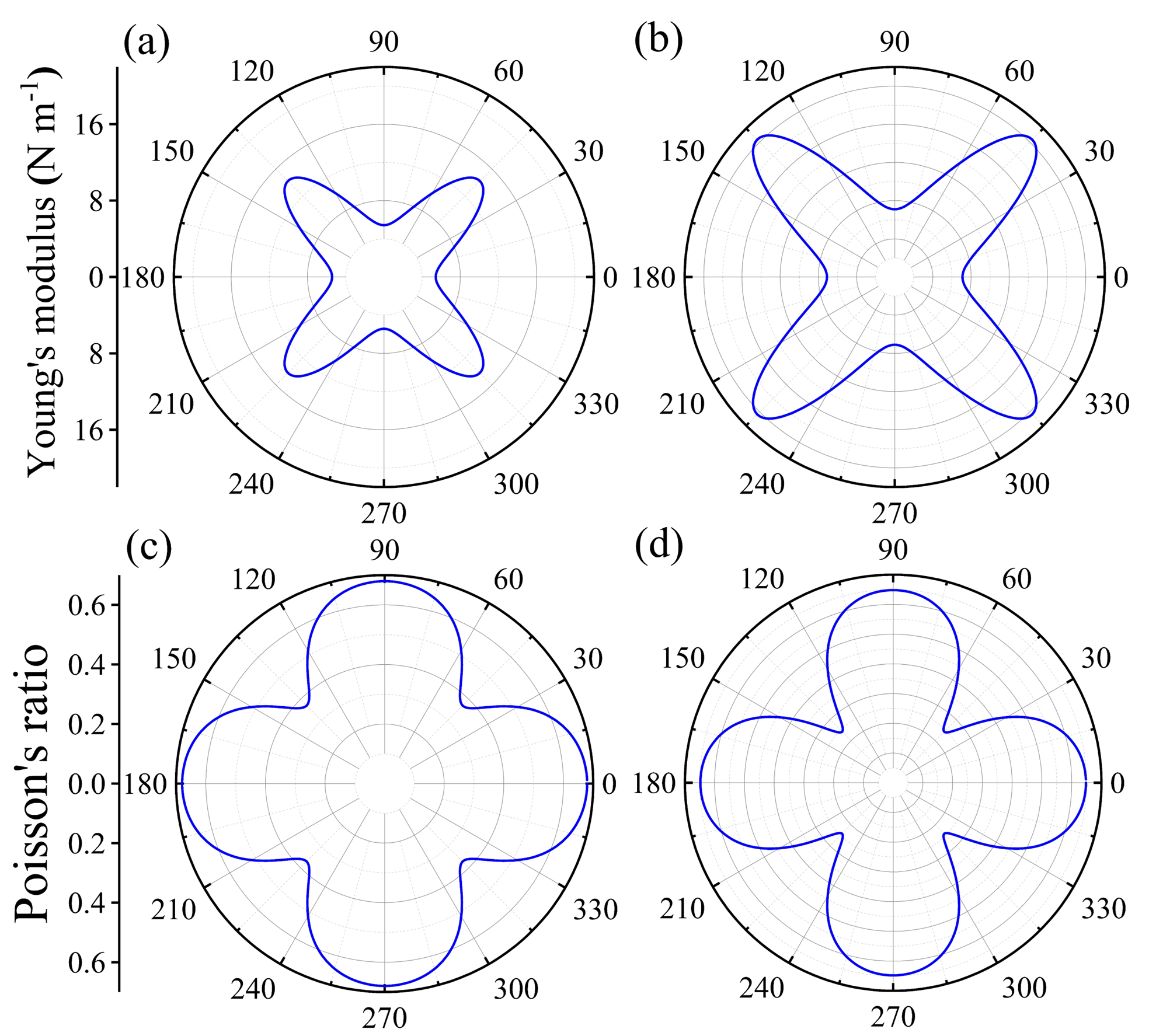}
\caption{(a)-(b) Young's modulus $Y(\theta)$ of Ti$_2$N$_2$ and Co$_2$N$_2$, respectively. (c)-(d) Poisson's ratio $\nu(\theta)$ of Ti$_2$N$_2$ and Co$_2$N$_2$,  respectively.}
\label{fig.3}
\end{figure}

\subsection{Mechanical properties}

The formulae for the layer modulus $\gamma$, 2D Young's moduli $Y$, Poisson's ratios $\nu$, and 2D shear modulus $G^{2D}$ are displayed below ~\cite{Andrew2012},
\begin{equation}
    {\gamma}=\frac{1}{4}({C}_{11}+{C}_{22}+2{C}_{12})
\end{equation}
\begin{equation}
    {Y}_{10}^{2D}=\frac{{C}_{11}{C}_{22}-{C}_{22}^{2}}{{C}_{22}} \quad and \quad {Y}_{01}^{2D}=\frac{{C}_{11}{C}_{22}-{C}_{22}^{2}}{{C}_{11}}
\end{equation}
\begin{equation}
    {\nu}_{10}^{2D}=\frac{{C}_{12}}{{C}_{22}} \quad and \quad {\nu}_{01}^{2D}=\frac{{C}_{12}}{{C}_{11} }
\end{equation}
\begin{equation}
    {G}^{2D}={C}_{66}.
\end{equation}
According them, we calculate some elastic moduli for \textit{M}$_2$N$_2$ (\textit{M} = Ti, Co) and listed the data in Table \ref{Table.2}.
For \textit{M}$_2$N$_2$ (\textit{M} = Ti, Co), the elastic constants ${C}_{11}$ and ${C}_{22}$ are equal, which is related to the equivalence of $a$ and $b$ axes in the two 2D materials. The ${C}_{11}$ of Co$_2$N$_2$ is larger than that of Ti$_2$N$_2$, which means that Co$_2$N$_2$ has superior flexibility. We also calculate the shear modulus and find that Co$_2$N$_2$ has a larger shear modulus, indicating Co$_2$N$_2$ is more resistant to shear strain. The Young's moduli reflect the ability of the material to resist deformation, and it's clear from Table \ref{Table.2} that this ability of Co$_2$N$_2$ is better than  Ti$_2$N$_2$. We also calculate the Poisson's ratios of the materials and the results show that all the two materials are tough materials.

\begin{table}[b] \centering
\caption{Layer modulus, 2D Young's moduli, Poisson's ratios, and 2D shear modulus of Ti$_2$N$_2$ and Co$_2$N$_2$.}
\renewcommand\tabcolsep{9pt}
\renewcommand\arraystretch{1.5}
\begin{tabular*}{8.5cm}{ccccc} \hline\hline
 Material     & $\gamma$ & $G^{2D}$ & $Y_{10}^{2D}=Y_{01}^{2D}$ & $\nu_{10}^{2D}=\nu_{01}^{2D}$ \\ \hline
Ti$_2$N$_2$ & 84.65          & 61.26             & 54.27              & 0.68             \\
Co$_2$N$_2$ & 100.54         & 102.79            & 70.33              & 0.65            \\ \hline\hline
\end{tabular*}
\label{Table.2}
\end{table}

In order to further investigate the mechanical response properties of 2D materials, the calculations of Young's modulus $Y(\theta)$ and Poisson's ratio $\nu(\theta)$ are also carried out. The curves in Fig. \ref{fig.3} represent the trajectories of Young's modulus $Y(\theta)$ and Poisson's ratio $\nu(\theta)$ with respect to $\theta$, which shows that the two structures are anisotropic. The Young's moduli reach their minimum values when the $\theta$ values are 0$^{\circ}$, 90$^{\circ}$, 180$^{\circ}$ and 270$^{\circ}$. Meanwhile, they gradually increase with the angle varying, reaching their maximum values at $\theta$ values of 45$^{\circ}$, 135$^{\circ}$, 225$^{\circ}$ and 315$^{\circ}$.
According to Fig. \ref{fig.3}, the Poisson's ratios of the two structures vary with $\theta$, and both of them reach the maximum values at 0$^{\circ}$, 90$^{\circ}$, 180$^{\circ}$ and 270$^{\circ}$.

\subsection{Electronic structure}
\subsubsection*{\rm\textbf{1.\quad Charge transfer}}

 Using the Bader charge analysis method~\cite{Henkelman2006}, we quantitatively compute the charge transfer between \textit{M} and N atom in the \textit{M}$_2$N$_2$ (\textit{M} = Ti, Co) layers. The charge transferred is 1.85 electrons from Ti to N atom, and 1.03 electrons from Co to N atom. The difference is due to the stronger metallicity of Ti atom than Co atom.

\subsubsection*{\rm\textbf{2.\quad Magnetic ground state}}

\begin{table*}[htbp] \centering
\caption{The energies of TiN and CoN in the FM, AFM-I, AFM-II phases, and the magnetic moments of individual Ti and Co atoms, and the exchange couplings \textit{J}$_1$, \textit{J}$_2$, magnetic anisotropic energy A and magnetic transition temperatures T$_C$/T$_N$ for \textit{M}$_2$N$_2$ (\textit{M} = Ti, Co). The energy of FM is set to 0 meV. The units of \textit{J}$_1$, \textit{J}$_2$, A are meV/S$^2$. The unit of T$_C$/T$_N$ is K.}
\renewcommand\tabcolsep{6.5pt}
\renewcommand\arraystretch{1.5}
\begin{tabular*}{17.8cm}{ccccccccccccc} \hline\hline
       & $ E_{FM}$   (meV) & $E_{AFM-I}$   (meV) & $E_{AFM-II}$   (meV) & Moment   ($\mu_B$) & \textit{J}$_1$ (meV/S$^2$) & \textit{J}$_2$ (meV/S$^2$) & A (meV/S$^2$) & T$_C$/T$_N$ (K) \\ \hline
TiN &  0     & 4.67          & 23.64          & 0.25  & -1.17        & -5.33        & -0.002    & 62    \\
CoN &  0    & -210.45              &  -643.86   & 1.98  & 26.31       & 67.33       & -0.121    & 474   \\ \hline\hline
\end{tabular*}
\label{TABLE.3}
\end{table*}

In the 3d transition-metal compounds, the electron-electron correlation is difficult to describe due to the complex \textit{d} electron interaction. In order to improve the accuracy of the electronic structure simulations of the \textit{M}$_2$N$_2$ layer, we take into account the electron correlation correction in our calculations through the GGA + \textit{U} method. GGA + \textit{U} method is usually used to deal with the electronic structure of transition metal compounds, in which the Hubbard \textit{U} parameter is introduced to account for the correlation of the onsite Coulomb interaction. By the self-consistent calculation using the linear response method~\cite{Cococcioni2005}, we determine the Hubbard \textit{U} values of 6.21 eV for Ti$_2$N$_2$ and 5.34 eV for Co$_2$N$_2$.

\begin{figure}[htbp]
\centering
\includegraphics[width=7.0cm]{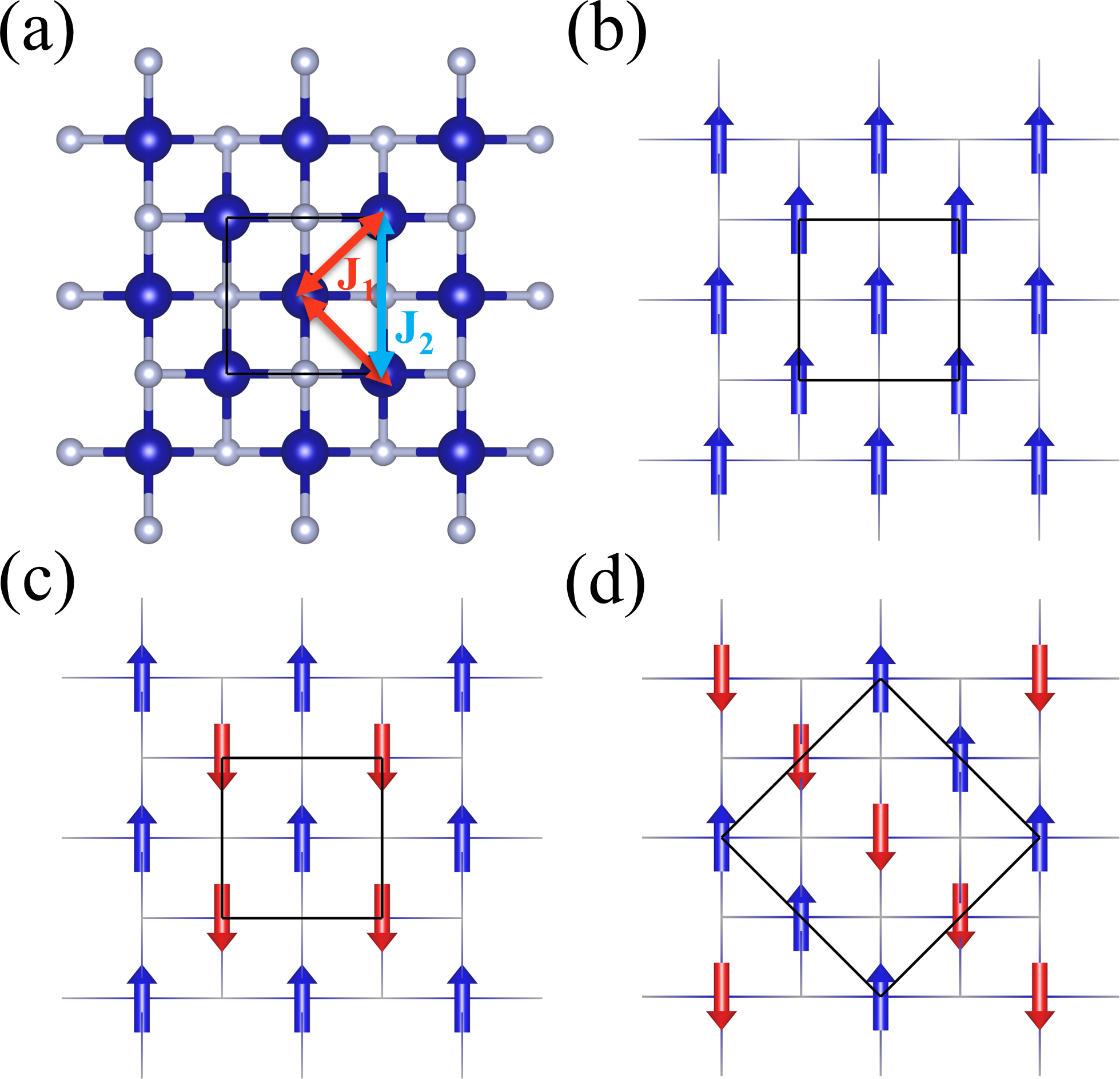}
\caption{(a) Top view of the atomic structure of \textit{M}$_2$N$_2$. \textit{J}$_1$ and \textit{J}$_2$ are the nearest and next-nearest neighbored exchange couplings. (b)-(d) The sketches of ferromagnetic order, antiferromagnetic order I and antiferromagnetic order II.}
\label{fig.4}
\end{figure}

\begin{figure}[htbp]
\centering
\includegraphics[width=8.5cm]{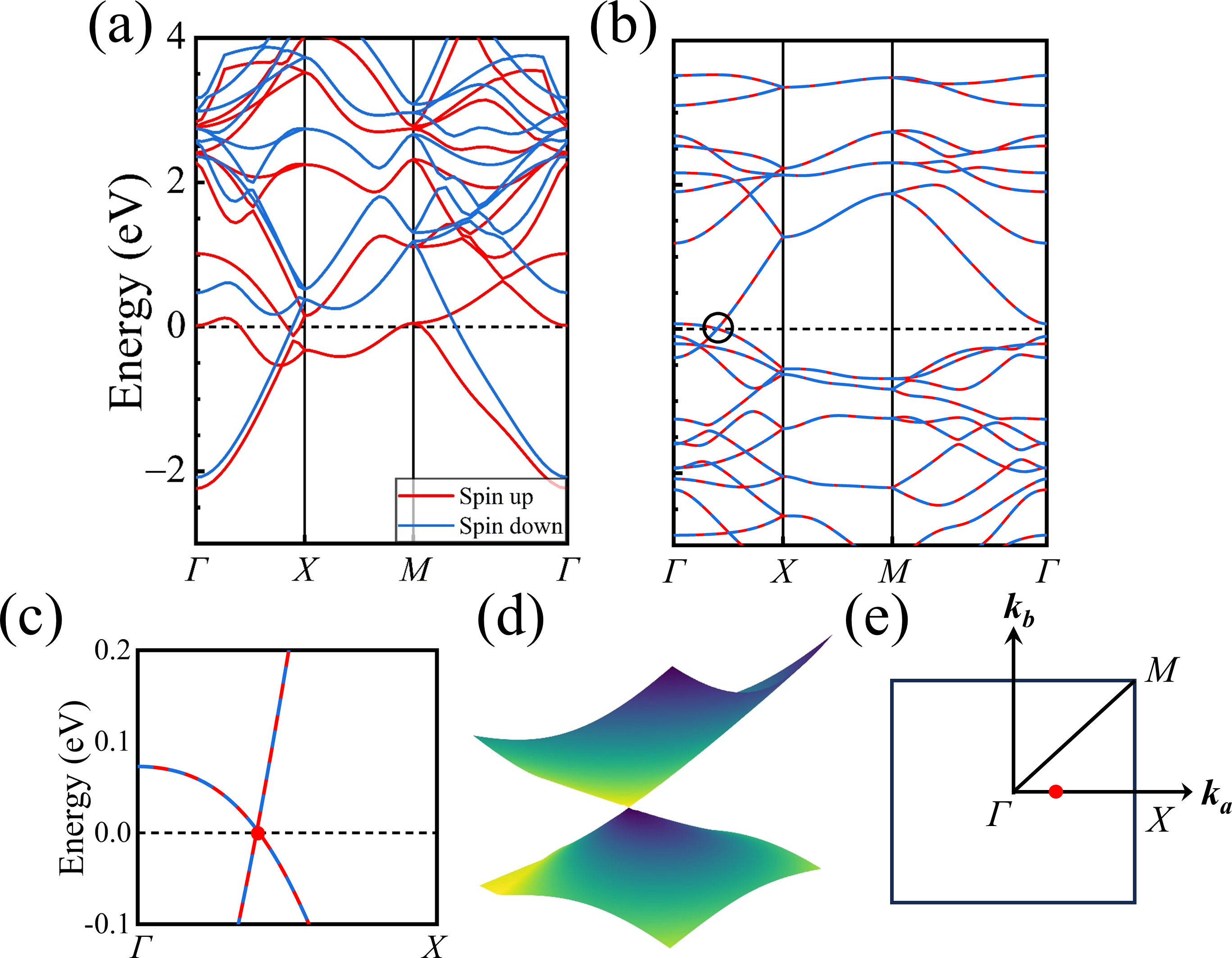}
\caption{(a)-(b) The band structures of Ti$_2$N$_2$ and Co$_2$N$_2$ in the ground states. The Dirac point is marked with a black circle. (c) The zoomed-in view of the Dirac point. (d) The three-dimensional energy dispersion of bands near the Dirac point. (e) The Brillouin zone of Co$_2$N$_2$ with high-symmetry k points labeled. The location of Dirac point is marked with a red dot.}
\label{fig.5}
\end{figure}

We study the magnetism of Ti$_2$N$_2$ and Co$_2$N$_2$ to clarify their magnetic ground states. Figure \ref{fig.4}(a) shows the top view of the atomic structure of \textit{M}$_2$N$_2$. The magnetic interactions of two neighbored \textit{M} atoms are marked with double arrow lines as well as \textit{J}$_1$ and \textit{J}$_2$. Figures \ref{fig.4}(b)-(d) show the ferromagnetic order (FM), antiferromagnetic order I (AFM-I) and antiferromagnetic order II (AFM-II). For the convenience of observing the magnetic order, the atomic structures are displayed with wire frames, in which the red and blue arrows represent two kinds of magnetic moments in opposite directions. The energies of Ti$_2$N$_2$ and Co$_2$N$_2$ in FM, AFM-I, and AFM-II order are computed afterwards. For comparison, we set the FM energy to 0 meV, and the energies relative to the FM energy are shown in Table \ref{TABLE.3}. The results indicate that Ti$_2$N$_2$ is a 2D ferromagnetic material and Co$_2$N$_2$ is a 2D antiferromagnetic material. The magnetic ground state of Co$_2$N$_2$ is AFM-II order. The single Ti and Co atoms have magnetic moments of 0.25 $\mu_B$ and 1.98 $\mu_B$, respectively.

\subsubsection*{\rm\textbf{3.\quad Band structure}}
Then, we calculate the band structures of Ti$_2$N$_2$ and Co$_2$N$_2$ in their magnetic ground states with the GGA + \textit{U} method, shown in Fig. \ref{fig.5}. The red and light blue curves correspond to the spin-up and spin-down bands. The highly symmetric points in reciprocal space are $\mathit{\Gamma}$(0 0 0), \textit{X}(0.5 0 0), and \textit{M}(0.5 0.5 0). As can be seen from Fig. \ref{fig.5}(a), the spin-up and spin-down bands have an obvious spin-splitting, and multiple bands cross its Fermi levels, indicating that the Ti$_2$N$_2$ is a magnetic metal.

In Fig. \ref{fig.5}(b), we find that an isolated Dirac point appears in the band structure of Co$_2$N$_2$, where two bands cross just at the Fermi level with linear dispersion. The zoomed-in graph is shown in Fig. \ref{fig.5}(c).
Besides, we plot the valence and conduction bands in the three-dimensional reciprocal space around the Dirac point, as shown in Fig. \ref{fig.5}(d).
The calculation indicates that the intersection of bands is indeed a Dirac point. Therefore, the Co$_2$N$_2$ monolayer is a 2D AFM Dirac semimetal with massless Dirac fermions.

\begin{figure*}[htbp]
\centering
\includegraphics[width=14cm]{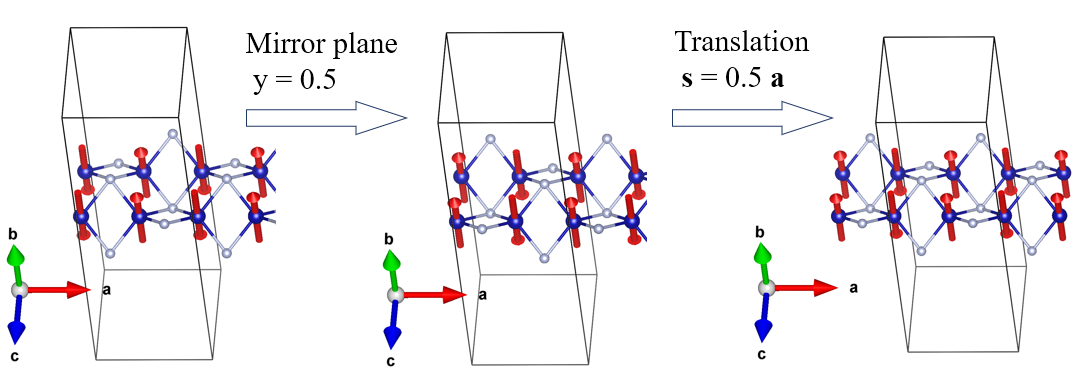}
\caption{The Mirror operation and translation operation in a glide mirror symmetry ${\tilde M}_y$, (x y z) $\rightarrow$ (x -y z+0.5). The mirror plane is $y$ = 0.5 and the translation vector is \textbf{s} = 0.5 \textbf{a}. The red arrows indicate the moment direction in the AFM-II order of Co$_2$N$2$ and the magnetic unit cell is marked with the solid line.}
\label{fig.6}
\end{figure*}

To deepen our understanding of the isolated Dirac point at Fermi level, we analyze the symmetry of the magnetic unit cell of Co$_2$N$_2$ monolayer.
In Fig. \ref{fig.6}, the red arrows parallel to $b$ axis indicate the direction of Co moments. The magnetic unit cell owns a glide mirror symmetry
${\tilde M}_y$, (x y z) $\rightarrow$ (x -y z+0.5). For clarity, the mirror operation and translation operation are shown in Fig. \ref{fig.6}, where the mirror plane is $y$ = 0.5 and the translation vector is \textbf{s} = 0.5 \textbf{a}.
As discussed in Ref. \citenum{Li2019c}, the Dirac point here is protected by the glide-mirror symmetry ${\tilde M}_y$. The two crossing bands have opposite ${\tilde M}_y$ eigenvalues along the $\Gamma$-X path, resulting in the linear crossing at the Dirac point.



\subsubsection*{\rm\textbf{4.\quad Electronic density of states}}
\begin{figure}[htbp]
\centering
\includegraphics[width=7.5cm]{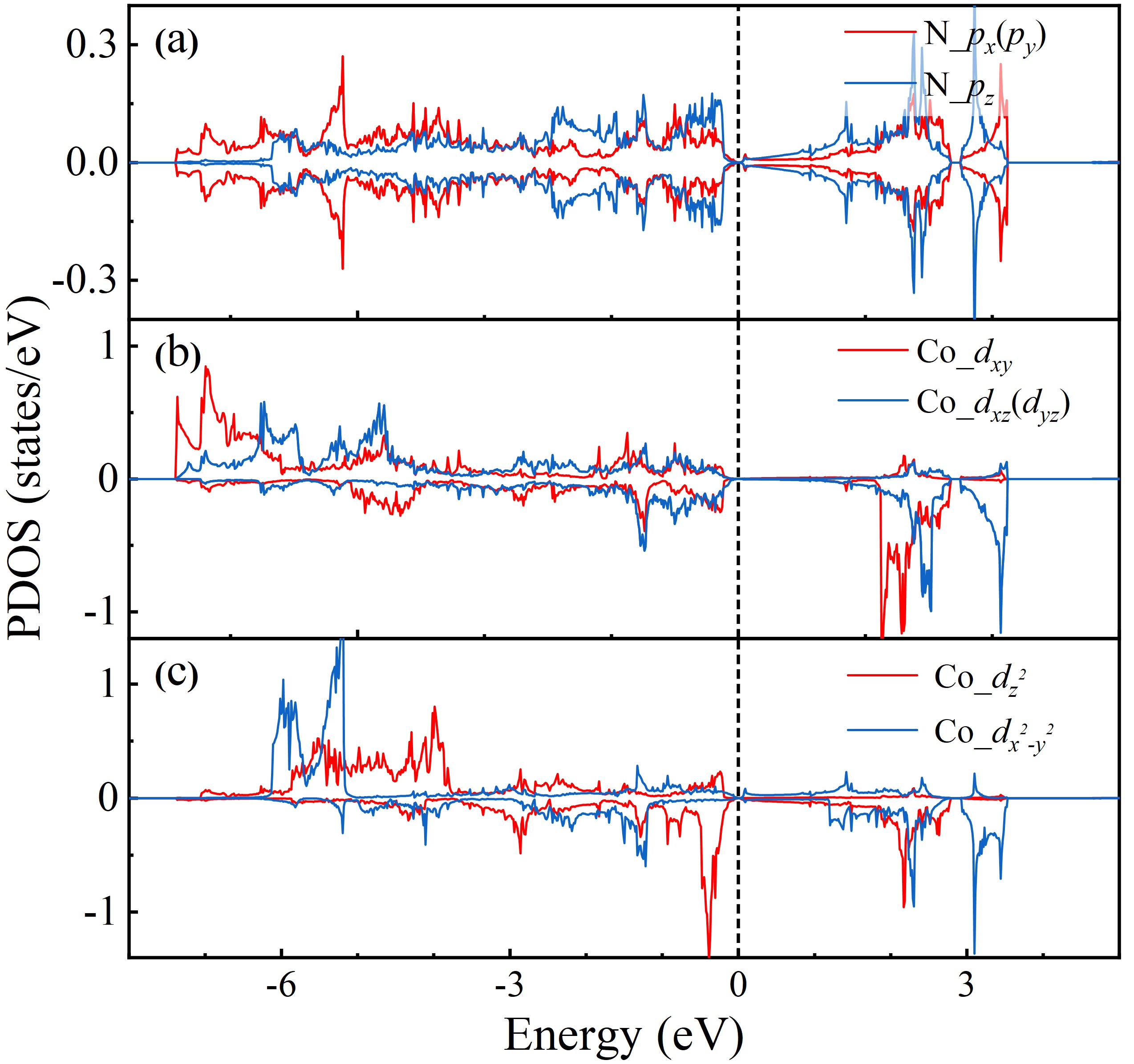}
\caption{Projected density of states(PDOS) on the Co \textit{3d} and N \textit{2p} suborbital. The Fermi energy represented by vertical dashed lines. The Fermi level is set to 0 eV.}
\label{fig.7}
\end{figure}

We take Co$_2$N$_2$ as the example to calculate the electronic density of its ground state. The PDOS of Co and N atoms is shown in Fig. \ref{fig.7}. The Co atom is at the center of the tetrahedron composed of four N atoms. According to the splitting of \textit{d} orbitals in tetrahedral crystal field, the five \textit{d} orbitals are divided into two groups: \textit{d$_{xy}$}, \textit{d$_{xz}$} and \textit{d$_{yz}$} belong to \textit{T$_{2g}$} orbital, \textit{d$_{x^2-y^2}$} and \textit{d$_{z^2}$} belong to \textit{E$_{g}$} orbital. In Fig. \ref{fig.7}(b), the PDOS of \textit{d$_{xy}$}, \textit{d$_{xz}$} and \textit{d$_{yz}$} are distributed in the same energy region. Particularly, \textit{d$_{xz}$} and \textit{d$_{yz}$} are completely degenerate. Since the Co-N tetrahedron is slightly compressed along the z axis, the \textit{d$_{xy}$} is not degenerate with the \textit{d$_{xz}$} and \textit{d$_{yz}$}. Similarly, the \textit{d$_{z^2}$} and \textit{d$_{x^2-y^2}$} in Fig. \ref{fig.7}(c) are also not degenerate due to tetrahedral distortion.

Fig. \ref{fig.7}(a) and (c) show that the PDOS of N \textit{p$_{x}$} and \textit{p$_{y}$} orbitals and Co \textit{d$_{x^2-y^2}$} orbital. The three orbitals posses the same shape and distribution and there is a strong orbital hybridization among them, which gives a major contribution to the formation of Co-N bonds.
 In addition, spin polarization calculation shows that a single Co atom has a magnetic moment of 1.98 $\mu_B$. In Fig. \ref{fig.7}(b), the three orbitals \textit{d$_{xy}$}, \textit{d$_{xz}$}, and \textit{d$_{yz}$} have stronger spin polarization than other orbitals, contributing 0.44, 0.43 and 0.43 $\mu_B$ to the magnetic moment of the Co atom, respectively.

\subsection{\rm\textbf{Magnetic transition temperature}}

Magnetic transition temperature is an important physical parameter of 2D magnetic materials, and we evaluate the Curie temperature of Ti$_2$N$_2$ as well as the N$\rm \acute{e}$el temperature of Co$_2$N$_2$ by the following methods. First, the Heisenberg model is used to describe the magnetic interactions in the \textit{M}$_2$N$_2$ (\textit{M} = Ti, Co) layers and the exchange couplings are calculated by the first principles method. The magnetic transition temperature is then determined by solving the Hamiltonian with Monte Carlo simulations~\cite{Liu2021b,Liu2022,Zhang2022}.

\begin{figure}[htbp]
\centering
\includegraphics[scale=0.32]{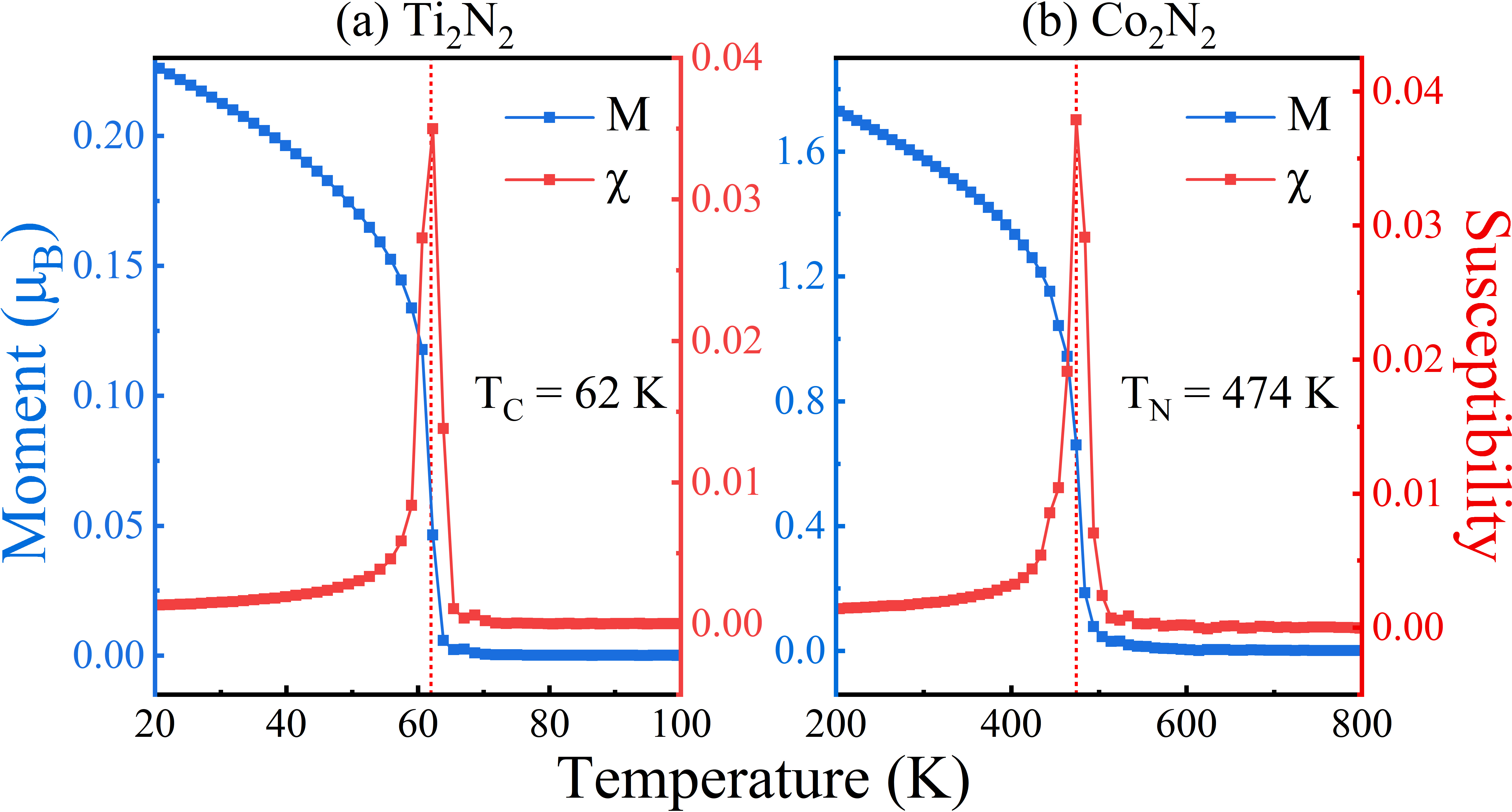}
\caption{The susceptibility ($\chi$) and average magnetic moment (\textit{M}) varying with temperature in \textit{M}$_2$N$_2$ (\textit{M} = Ti, Co). (a) Ti$_2$N$_2$, (b) Co$_2$N$_2$.}
\label{fig.8}
\end{figure}

The spin Hamiltonian in the 2D square lattice of \textit{M}$_2$N$_2$ monolayer is defined as
\begin{equation}
    H={J}_{1}\displaystyle\sum_{<ij>}\overrightarrow{S}_{i}\cdot\overrightarrow{S}_{j} +{J}_{2}\displaystyle\sum_{<<ij^{\prime}>>}\overrightarrow{S}_{i}\cdot\overrightarrow{S}_{j^{\prime}}+A\displaystyle\sum_{i}(S_{iz})^2
\end{equation}
where \textit{j} and \textit{j$^{\prime}$} denote the nearest and next-nearest neighbors of the \textit{i} site. Here, \textit{J}$_1$ and \textit{J}$_2$ are the nearest and next-nearest neighbored exchange couplings, A is the single-site magnetic anisotropic energy.
The exchange couplings \textit{J}$_1$ and \textit{J}$_2$ are defined as
\begin{equation}\label{J1J2}
\begin{aligned}
&{J}_{1}=({E}_{FM}-{E}_{AFM-I})/4,\\ &{J}_{2}=({E}_{FM}+{E}_{AFM-I}-2{E}_{AFM-II})/8,
\end{aligned}
\end{equation}
the values of \textit{J}$_1$ and \textit{J}$_2$ are derived from the energy differences among the FM order, AFM-I order and AFM-II order, listed in Table \ref{TABLE.3}.
 The variation of magnetic moment ($M$) and magnetic susceptibility ($\chi$ = $\frac{<\vec{M}^2> - <\vec{M}>^2}{k_BT}$) with temperature in Fig. \ref{fig.8}, which determine the N$\rm \acute{e}$el temperature of \textit{M}$_2$N$_2$ (\textit{M} = Ti, Co) to be  62 K and 474 K, respectively.


\section{CONCLUSIONS}

In summary, based on the first-principles calculations within the framework of the density functional theory, we design two 2D transition-metal nitrides \textit{M}$_2$N$_2$ (\textit{M} = Ti, Co), and their structural stability is demonstrated by the cohesive energy, formation energy, elastic constants, phonon spectra and molecular dynamics simulations. The mechanical properties of the two structures are anisotropic as illustrated by the elastic modulus values. Ti$_2$N$_2$ is a 2D ferromagnetic material, while Co$_2$N$_2$ is a 2D antiferromagnetic material with Dirac point at the Fermi level. In addition, the Heisenberg model with the Monte Carlo method is employed to determine the magnetic transition temperature, and the N$\rm \acute{e}$el temperature in the Co$_2$N$_2$ layer is found to be as high as 474 K. Therefore, the Co$_2$N$_2$ monolayer is a rare antiferromagnetic 2D material with both high temperature and Dirac points.

\begin{acknowledgements}
This research was funded by the National Natural Science Foundation of China under Grants Nos. 12274255, 12274458, 11974207.
\end{acknowledgements}

\bibliography{library}

\end{document}